\documentclass{aastex63}
\usepackage[utf8]{inputenc}
\usepackage{makecell}

\usepackage{graphicx}
\usepackage{hyperref}

\def\flu{{\cal F}}
\def\flucrit{\flu^{\rm crit}}
\def\lethdist{D^{\rm leth}}
\def\pfrac#1#2{\left( \frac{#1}{#2} \right)}

\def\iso#1#2{\mbox{${}^{#2}{\rm #1}$}}
\def\fe6#1{\iso{Fe}{6#1}}
\def\pu24#1{\iso{Pu}{24#1}}

\begin{document}

\title{X-ray-luminous Supernovae: Threats to Terrestrial Biospheres}
\author[0000-0002-2478-3084]{Ian R. Brunton}
\affiliation{Department of Astronomy, University of Illinois Urbana-Champaign, Urbana, IL 61801}
\affiliation{Astromaterials Research \& Exploration Science, NASA Johnson Space Center, Houston, TX 77058}
\affiliation{Illinois Center for Advanced Studies of the Universe, University of Illinois Urbana-Champaign, Urbana, IL 61801}
\author[0000-0001-8315-7385]{Connor O'Mahoney}
\affiliation{Department of Astronomy, University of Illinois Urbana-Champaign, Urbana, IL 61801}
\affiliation{Illinois Center for Advanced Studies of the Universe, University of Illinois Urbana-Champaign, Urbana, IL 61801}
\author[0000-0002-4188-7141]{Brian D. Fields}
\affiliation{Department of Astronomy, University of Illinois Urbana-Champaign, Urbana, IL 61801}
\affiliation{Illinois Center for Advanced Studies of the Universe, University of Illinois Urbana-Champaign, Urbana, IL 61801}
\affiliation{Department of Physics, University of Illinois Urbana-Champaign, Urbana, IL 61801}
\author[0000-0001-9632-8307]{Adrian L. Melott}
\affiliation{Department of Physics and Astronomy, University of Kansas, Lawrence, KS 66045}
\author[0000-0001-9091-0830]{Brian C. Thomas}
\affiliation{Department of Physics and Astronomy, Washburn University, Topeka, KS 66621}

\begin{abstract}
The spectacular outbursts of energy associated with supernovae (SNe) have long motivated research into their potentially hazardous effects on Earth and analogous environments. Much of this research has focused primarily on the atmospheric damage associated with the prompt arrival of ionizing photons within days or months of the initial outburst, and the high-energy cosmic rays that arrive thousands of years after the explosion. In this study, we turn the focus to persistent X-ray emission, arising in certain SNe that have interactions with a dense circumstellar medium and observed months and/or years after the initial outburst. The sustained high X-ray luminosity leads to large doses of ionizing radiation out to formidable distances. We assess the threat posed by these X-ray-luminous SNe for Earth-like planetary atmospheres; our results are rooted in the X-ray SN observations from Chandra, Swift-XRT, XMM-Newton, NuSTAR, and others. We find that this threat is particularly acute for SNe showing evidence of strong circumstellar interaction, such as Type IIn explosions, which have significantly larger ranges of influence than previously expected and lethal consequences up to $\sim 50$ pc away. Furthermore, X-ray-bright SNe could pose a substantial and distinct threat to terrestrial biospheres and tighten the Galactic habitable zone. We urge follow-up X-ray observations of interacting SNe for months and years after the explosion to shed light on the physical nature and full-time evolution of the emission and to clarify the danger that these events pose for life in our galaxy and other star-forming regions.

\end{abstract}

\section{Introduction}

The spectacular outbursts of energy originating from SNe have long motivated research into the harmful effects they may impose on Earth and analogous environments. The early work of \citet{schindewolf1954} and \citet{Krass58} promptly recognized the importance of ionizing radiation on Earth’s atmosphere and biosphere, which has been the central focus of subsequent work \citep[][and references below]{TerryTuck68, Ruder74, whitetal1976, elschramm1995, Gehrels2003}. More generally, these events constrain the Galactic habitable zone, i.e., the locations throughout the galaxy in which life could exist \citep{Line2004, Gowan2011,Cockell2016}. 

Naturally, the field of nearby SN research develops alongside our understanding of SNe in general. \citet{Gehrels2003}, for example, specifically evaluated the near-Earth SN threat in light of multiwavelength observations and theoretical models of the lone event of SN 1987A. But since then, a more profound understanding of SN characteristics has developed and so too has our insight into their influence on terrestrial atmospheres and habitability. In this paper, we now examine the consequences of SN X-ray emission stemming from observations largely made in the years following the Gehrels study. 

Further motivating our research is that there is now a wealth of empirical evidence for near-Earth SNe in the geologically recent past. The radioactive isotope \fe60 (half-life 2.6 Myr) has been found live (not decayed) in deep-ocean samples dating 2--3 Myr ago \citep[ferromanganese crusts;][]{Knie99, Knie04, Fits2008, Wallner16, Ludwig16, Wall21}. \fe60 is also found in Apollo samples of lunar regolith \citep{Fimiani2016}, in cosmic rays \citep{Binns2016}, and (with a smaller flux) in recent deep-ocean sediments \citep{Wallner2020} and modern Antarctic snow \citep{Koll2019}. The crust and sediment measurements all indicate an event occurred around 3 Myr ago, and \citet{Wall21} now find evidence for another event around 7--8 Myr ago.

The widespread geological presence of pulses of live SN-produced radioiostopes is the hallmark of near-Earth events \citep{EllisFieldsSchramm1996}. Furthermore, the \fe60 abundances for these events allow estimates of the SN distance, around 20--150 pc \citep{FieldsEllis1999, FieldsEllisHochmuth2005, Fry_2015}, and candidate star clusters have been proposed at distances around 50 pc and 100 pc \citep{Benitez2002,Mamajek2007,Hyde2018}. Indeed, close-by events are required to successfully deliver SN ejecta to Earth \citep{FieldsAnthanassiadouJohnson2008, Fry_2015, Fry_2016}. In addition, the \citet{Wall21} discovery of \pu244 in the same time window further strengthens the case for nearby explosions, and opens the possibility of an additional event from a kilonova \citep{Wang2021}. These detections are entirely consistent with the presence of our solar system within the Local Bubble, a hot, low-density region of space that is thought to be a product of numerous nearby SN explosions coinciding with Earth’s early Neogene Period ($\lesssim$ 20 Myr) \citep{Breit2016, Zucker22}. Altogether, these observations demonstrate that nearby explosions are a fact of life in our star-forming galaxy and suggest that even closer events are possible over the history of biology on Earth, possibly causing mass extinctions \citep{Fields2020}.

\subsection{General Terrestrial Effects of a Nearby Supernova}

Each SN will have its own unique characteristics and evolutionary behavior, with variables resulting from both the internal and external environment of the progenitor star \citep{alsmurd2017}. Most important to the effects imposed on a nearby planet in the interstellar neighborhood, however, are the flux, spectrum, and duration of the ionizing radiation emitted as a result of the event. The exact perturbations that are then imposed on the planetary system will be dependent on the characteristics of the planet itself. We are primarily concerned here with the threats to terrestrial biospheres. As such, with Earth still being the sole confirmed substrate for any sort of viable life-form, we typically situate our analysis of a nearby SN with respect to the deleterious effects that would pertain to Earth’s modern environment.

In general, for a terrestrial planet that harbors a robust atmosphere, the most direct effects will be triggered by the radiative alteration of the planet’s atmospheric chemistry. The multifarious physical processes governing such interactions can make it difficult to parameterize the exact implications. But, early research as far back as \citet{Krass58} speculated on the general implications that nearby SNe may have on life, and numerous others have researched the habitability consequences of high influxes of radiation from a variety of interstellar events \citep[][to name a very select few]{Mart2011,Chen2018,Louca2022,Amb2022}.

With such a complex system as Earth’s atmosphere, any large, energetic perturbation would induce a wide range of responses. Moreover, merely assessing the threat imposed on an Earth-like biosphere must take into account the multiple different phases of Earth’s atmosphere and even the nature of life throughout its 4.5 Gyr history. The effects on the biosphere, say, during the Archean Earth (\textgreater 2.5 Gya), prior to the oxygenation of the atmosphere or land-based life-forms, would be entirely different from effects imposed on the oxidizing atmosphere and complex organisms of the current Phanerozoic Eon (\textless 540 Mya).

Consequences related to atmospheric heating and escape are a focus of much habitability work related to other energetic events, such as outflows from active galactic nuclei \citep[AGN;][]{Amb2022}, and extreme ultraviolet radiation from quasars \citep{Chen2018}. But while the X-ray emission from SNe are indeed high-energy phenomena, the total, time-integrated energy output of radiation pales in comparison to those more sustained events, and the SN would need to be at such close distances to impose any significant heating or escape that other factors would dominate at that point. Notably, however, \citet{Smith2004} calculated that for a hypothetical biosphere located on a planet with a far thinner atmosphere than found on Earth, even atmospheric heating and escape effects may be significant for an SN. Speculations about hypothetical biospheres are beyond the scope of this paper.

\citet{Smith2004} also analyzed the propagation of ionizing radiation through model atmospheres of terrestrial exoplanets. They found that while even the thinnest of atmospheres will often block all X-ray radiation from reaching the surface, there may be a significant amount of incident X-ray energy that will redistribute into diffuse, but “biologically effective,” UV at the surface. More generally, the incident X-ray energy can produce substantial transient fluctuations in atmospheric ionization levels — an effect that would be particularly relevant for advanced, technological civilizations (e.g., radio communication). These sudden ionospheric disturbances are certain to occur in the event of an influx of X-ray radiation, and are often a major focus of research regarding the influx of X-rays from large solar flares \citep{Mitra74,Hayes2017,Hayes2021,Sisk2022}.

While these cited studies have examined the atmospheric influences of ionizing radiation, a notable barrier to extrapolating directly from these previous data (and many others, such as \citet{Seg10, Chen21}) is that none yet have resolved the particular influences of the X-ray phase of SN emission. For example, the \citet{Smith2004} study utilized spectral models most relevant for solar flares, gamma-ray bursts (GRBs), and the non-X-ray-specific emission of SNe. The emission profiles of the X-ray-luminous SNe are far more energetic, enduring, and of harder photon spectra than those of solar flares, but particularly less energetic and of softer photon spectra than GRBs. This is a commonality among all the modeled results we reference herein since this is the first paper we are aware of that is specific to the distinct, X-ray phase of SN emission. We intend to more accurately model the X-ray emission profile against a climate model in a future study.

\subsection{Ozone-related Effects of a Nearby Supernova}

Since our research is the first to focus solely on the X-ray phase of SN emission, we elect to provide a broad assessment of the lethal effects that sits within the context of most prior research into SN threats to habitability. As such, we focus the majority of our analysis only on the deleterious mechanism of ozone depletion. Much previous literature on the habitable influence of SNe often addresses the loss of stratospheric ozone as the most notable of the general consequences to Earth’s modern-day atmosphere, as it is the ozone that currently serves as the biosphere’s most formidable line of defense against external radiation \citep{whitetal1976,elschramm1995,Gehrels2003,ThomMel2005a,ThomMel2005b,Ejzak2007,Melott2011}. Justification for this approach with respect to hard X-rays is provided further in \S \ref{sect:meth}.

\citet{Ruder74} was likely the first to orient focus on the catalytic cycle of ozone depletion that would result from photoionization delivered by the SN explosion. Subsequent research into the effects of nearby SNe on habitability has thus similarly used ozone loss as a proxy for biological damage, and hence how “lethal” an SN will be \citep*[see][for a detailed review]{Melott2011}.

This prior research often pertains to the two distinct phases of ionizing radiation that are present in all SNe regardless of environment: (1) the prompt arrival of energetic photons — predominantly gamma rays — associated with the massive outbursts of energy, and (2) the later influx of charged cosmic rays that propagate outwards with the SN blast. 

Despite the magnificence of the initial outburst, the accompanying gamma rays radiate $E_\gamma \sim 2 \times 10^{47} \ \rm erg$ \citep{Gehrels2003}, a small fraction of the $\sim 10^{51} \ \rm erg$ of the blast energy. As we will see below, for likely SN distances, this emission is unable to trigger a notable rate of ozone depletion unless further enhanced by subsequent X-ray emission and, therefore, does not represent a significant threat to life on its own. The exception to this occurs in the unique circumstances of an SN accompanied by a GRB, where relativistic jets lead to beamed gamma-ray emission. Though astronomically rare, GRBs from SNe and compact object mergers are hypothesized to have also directly influenced Earth’s geological past \citep{ScaloWheel02, Melott2004, ThomMel2005a, ThomMel2005b, Piran2014}. In fact, the recent GRB 2210009A, which was over 700 Mpc away from Earth \citep{de2022,Will23}, actually caused a slight, but measurable, perturbation in the \emph{D}-region of the ionosphere \citep{Hayes2022}; fortunately, this delivered far below any type of lethal dosage.

A later, more destructive phase of SN ionizing radiation occurs hundreds to tens of thousands of years after the initial arrival of photons, in which a planet’s atmosphere is bathed with an influx of cosmic rays freshly accelerated by the SN. Cosmic rays carry a much larger fraction of the SN blast energy ($\ga 10\%$) nd so present a greater threat. The time history of SN cosmic rays experienced on Earth depends on their distribution in and around the SN remnant that accelerates them. The escape of cosmic rays from SN remnants remains a topic of active research and depends on the cosmic-ray energy and remnant age. For example, the highest-energy cosmic rays can escape early to act as precursors to the SN blast, while lower-energy particles that are the bulk of cosmic rays are advected into the remnant and remain there until its end stages due to self-confining magnetohydrodynamic instabilities excited by the cosmic rays themselves \citep{Drury2011, Nava2016}. Broadly speaking, for an observer on Earth, the cosmic flux should be the most intense around the time the forward shock arrives, which is thousands of years after the explosion. 

The spectrum of the newly accelerated cosmic rays should have more high-energy particles than the equilibrium propagated spectrum throughout most of the galaxy \citep{Aharonian1996,Telezhinsky2012,Bell2013,Brose2020}. And with the bulk of the cosmic rays trapped in the remnant, the duration of the SN cosmic-ray exposure will be that of the blast passage — many thousands of years.

If the SN is close enough that the biosphere is exposed to a dose of cosmic rays, this can represent the most harmful stage of a nearby SN on a terrestrial environment, simply due to the extremely high energy and prolonged influential presence of the cosmic rays. As these cosmic rays penetrate deep into the atmosphere, lingering for hundreds to thousands of years, they alter the atmospheric chemistry of the planet. It is this phase that has typically been the focus of prior research on the terrestrial consequences of SNe. In said research, it is generally assumed that a rate of $30\%-50\%$ globally averaged ozone depletion (in reference to modern-day Earth levels) harbors the potential to impose an extinction-level event on Earth \citep{Melott2011}. The distance the SN would have to be to impose such consequences is often labeled the “lethal distance.” In their review of astrophysical ionizing radiation and Earth, \citet{Melott2011} have surmised a typical lethal distance for an SN to be around 10 pc. However, this value is largely variable due to a variety of factors, and in a more recent detailed study, \citet{melthom2017} have also estimated that this distance may be upwards of 50 pc for certain interstellar conditions. These general parameters serve as the foundation for our assessment of the threat imposed by X-ray-luminous SNe.

\subsection{X-Ray-luminous Supernovae}
\label{sect:x-raySN}

\begin{figure}
    \centering
    \includegraphics[width=0.8\textwidth]{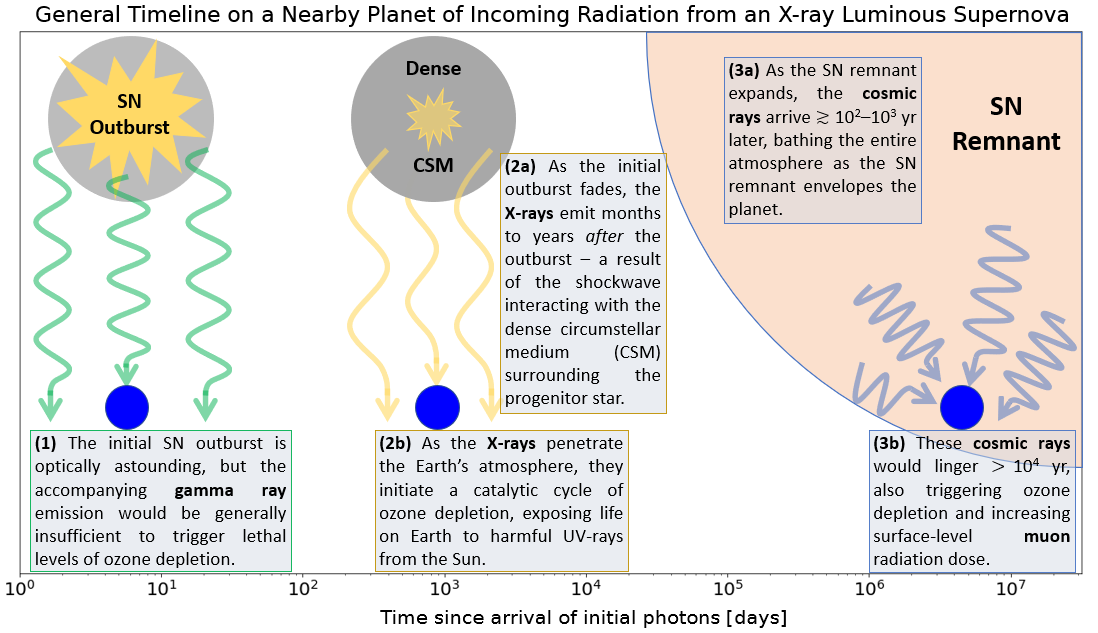}
    \caption{Hypothetical timeline on a planet for the arrival of radiation emitted by a nearby X-ray-luminous SN. Note that the timeline is logged, displayed in days since the initial outburst is seen in the sky. The three phases from left to right: (1) The initial arrival of gamma rays and other photons (green) in the SN outburst. (2) The X-ray phase of emission (yellow) delayed by months to years after the outburst. (3) The arrival of cosmic rays (blue) with the SN remnant thousands of years later.}
    \label{fig:SN Evolution}
\end{figure}

In this paper, we now turn the analytical focus to the persistent high-energy X-ray emission that characterizes certain SNe. The timing of the X-ray emission is situated in between the two previously mentioned general stages: months and/or years after the initial outburst and hundreds to thousands of years before the arrival of cosmic rays. Thus, the associated damage from the X-ray influx would occur as a separate phase of ionizing radiation, altering our notions of the timeline by which an SN influences a nearby planet.

Figure \ref{fig:SN Evolution} illustrates a hypothetical, but realistic, timeline by which an X-ray-luminous SN’s radiation would interact with a nearby planet (note that the timeline uses a log scale). The three phases from left to right are (1) the initial arrival of photons (green) in the SN outburst, (2) the X-ray phase of emission (yellow) delayed by months to years after the outburst, followed by (3) the arrival of cosmic rays (blue) with the SN remnant thousands of years later. 

The X-ray emission is thought to arise from interactions with a dense circumstellar medium (CSM) carved out during the star’s lifetime \citep{Smith2014,Chandra18,Dwark2019}. It has long been theorized \citep{Ruder74} that some SNe may have high-enough X-ray emission to trigger the catalytic cycle of ozone depletion similar to that of the later cosmic rays — albeit at different magnitudes and shorter timescales. And while X-ray significance has been acknowledged from the beginning, the absence of empirical X-ray SN observations to parameterize such discussion has correlated with their absence from discussions on planetary influence.

In the last few decades, the field of X-ray astronomy has come of age \citep{Wilkes22}, most notably due to innovations in telescopic sensitivity employed through Chandra, Swift-XRT, XMM-Newton, and NuSTAR. The combined capabilities of these telescopes have allowed us to gain a keen understanding of the X-ray evolution from all types of SNe, and these observations have served to both confirm and upend certain notions of stellar processes \citep{Dwark2014,Wilkes22}. 

For this study, we have scoured the literature for X-ray SNe to compare the evolutionary characteristics of 31 X-ray-luminous SNe (occasionally called “interacting SNe,” due to the necessary interaction with a CSM). We use these data to conduct a threat assessment that serves as an initial parameterization for the impact that SN X-ray emission can have on terrestrial planets. In \S \ref{sect:meth}, we briefly summarize the general threat to terrestrial planets that is imposed by ionizing photon events, explaining the key value of critical fluence and determining an appropriate fluence value for the associated X-ray data. This discussion outlines the process by which we can then parameterize our subsequent assessment of the X-ray emission.

In \S \ref{sect:data} we then present the collection of light curves for the 31 SNe analyzed. We identify the key characteristics of each spectral type and identify the emission trends relevant to terrestrial atmospheres. Throughout our analysis of the data, we refrain from including values based on speculation of any unconfirmed X-ray emission, i.e., we only display confirmed X-ray observations recorded in the literature. However, as will be discussed in \S \ref{sect:data}, there are clear indications that each SN actually has a larger amount of total X-ray emission than has been observed. Therefore, by limiting our data analysis to the confirmed observations only, we are providing a conservative, lower-end estimate of their overall X-ray energy output. Nevertheless, we maintain this restriction, because an overarching purpose of this paper is to provide empirical evidence for the threat that SN X-ray emission can impose on terrestrial planets. This subsequently enables us to provide a conservative estimate for the ranges at which the X-ray-luminous SNe would be threatening to a nearby biosphere, and thus, when we present our threat assessment in \S \ref{sect:res}, we will have gone to great lengths to show that the inferences we make are appropriate, and the generalizing rate of lethality for X-ray-luminous SNe is significant.

We discuss our results in \S \ref{sect:disc} and conclude with the summation of our findings in \S \ref{sect:summ_and_conc}, most notably, that SN X-ray emission — a distinct phase of an SN’s evolution — can certainly impose effects on terrestrial atmospheres and biospheres at formidable distances. These results have implications for planetary habitability, the Galactic habitable zone, and even Earth’s own evolutionary history.

\section{Methods: Parameterizing X-Rays and Ozone Depletion}
\label{sect:meth}

\begin{table}
\begin{center}
\begin{tabular}{|l l l l|} 
 \hline
 Supernova & Publication(s) & Band Observed (keV) & \\ [0.5ex] 
 \hline
 1970G & \citet{2005ApJ...632L..99I}  & 0.3--2.0 &\\ 
 1978K & \citet{2004ApJ...603..644S} & 0.5--2.0 &\\
 1979C & \citet{2011NewA...16..187P}  & 0.3--2.0 &\\ 
 1980K & \citet{1995RPPh...58.1375S} & 0.2--2.4 &\\
 1986J & \citet{2005MNRAS.362..581T}  & 0.3--2.0 &\\ 
 1987A & \citet{2016ApJ...829...40F} & 3.0--8.0 &\\
 1988Z & \citet{2006ApJ...646..378S}  & 0.2--2.0 &\\ 
 1993J & \citet{2009ApJ...699..388C} & 0.3--8.0 &\\
 1994I & \citet{2002ApJ...573L..27I}  & 0.3--2.0 &\\ 
 1995N & \citet{2000MNRAS.319.1154F, 2005ApJ...629..933C} & 0.1--10; 0.1--10 &\\
 1996cr & \citet{2008ApJ...688.1210B}  & 2.0--8.0 &\\ 
 1998S & \citet{2002ApJ...572..932P} & 2.0--10 &\\
 2001em & \citet{Chandra20} & 0.3--10 &\\
 2003bg & \citet{2006ApJ...651.1005S} & 0.3--10 &\\
 2004dj & \citet{2012ApJ...761..100C} & 0.5--8.0 &\\
 2004dk & \citet{2019ApJ...883..120P} & 0.4--8.0 &\\
 2004et & \citet{2007MNRAS.381..280M} & 0.5--8.0 &\\
 2005ip & \citet{2014ApJ...780..184K,2017MNRAS.466.3021S,2020MNRAS.498..517F} & 0.2--10; 0.5--8.0; 0.5--8.0 &\\
 2005kd & \citet{2016MNRAS.462.1101D, 2016ApJ...832..194K} & 0.3--8.0; 0.2--10 &\\
 2006bp & \citet{2007ApJ...664..435I} & 0.2--10 &\\
 2006jc & \citet{2008ApJ...674L..85I} & 0.2--10 &\\
 2006jd & \citet{2012ApJ...755..110C, 2016ApJ...832..194K} & 0.2--10; 0.2--10 &\\
 2008D & \citet{mod2009} & 0.3--10 &\\
 2010jl & \citet{2015ApJ...810...32C} & 0.2--10 &\\
 2011ja & \citet{2013ApJ...774...30C} & 0.3--10 &\\
 2012ca & \citet{Boch18} & 0.5--7.0 &\\
 2013by & \citet{2013ATel.5106....1M, 2017ApJ...848....5B} & 0.3--10; 0.3--10 &\\
 2013ej & \citet{2016ApJ...817...22C} & 0.5--8.0 &\\
 2014C & \citet{Bret2022} & 0.3--100 &\\
 2017eaw & \citet{2019ApJ...876...19S} & 0.3--10 &\\
 2019ehk & \citet{2020ApJ...898..166J} & 0.3--10 &\\
 \hline
\end{tabular}
\end{center}
\caption{The list of 31 SNe for which multiple data points of X-ray luminosities were found within the literature. Also cited are the original publication(s) from which we extracted the observations, along with the corresponding energy band reported for each observation. Most notable is that the range of cited observations used in this paper are all $\leq$ 10 keV photon energy levels (SN 2014C excepted). We account for this in our analysis of the fluence required for certain levels of ozone depletion.}
\label{tab:SNproperties}
\end{table}

A multitude of physical processes would occur in response to the high influx of X-ray radiation from an SN. But due to the overwhelming and persistent emission spectra associated with the X-ray profile, the dominant effect of this influx on modern-day Earth is the alteration of the planet’s atmospheric chemistry. Most significant for the biosphere (and thus lethality) is the radiative breaking of the chemical bond of N$_2$. Once this bond breaks, the nitrogen interacts with neighboring atmospheric oxygen and prompts the generation of nitrogen oxides. These nitrogen compounds (often identified NO$_x$) catalyze a cycle of ozone depletion by converting O$_3$ to O$_2$ \citep[see][for a more detailed description of this process]{Ruder74,Solo1982, Gehrels2003, Rohen2005,Melott2011}. At very high altitudes, a similar mechanism known as HO$_x$-induced ozone depletion would dominate, but it would not be particularly relevant for lethality due to the very short lifetime of HO$_x$ constituents \citep{Rohen2005,Solo1983}.

On Earth, stratospheric ozone typically acts as the primary absorber of UVB radiation from the Sun, radiation that is extremely damaging to most organisms. As such, a significant loss of ozone would expose life on Earth to high doses of UVB radiation. Changes in irradiance at Earth’s surface and into the ocean have variable effects across the biosphere \citep{Thomas2015, Neale2016, Thom2018} with UVB negatively influencing a wide range of organisms, especially marine organisms at the lower end of the food chain. It has been suggested that such effects may be significant enough to initiate a mass extinction event \citep{elschramm1995, Melott2004, melthom09, Beech2011, Piran2014, Fields2020}.

Previous studies assessing the lethality of SNe have thus used ozone depletion as the predominant proxy for biological damage \citep{Ruder74, whitetal1976, Gehrels2003, Ejzak2007}. For the purposes of comparison, we follow this procedure, noting that there will of course be other atmospheric effects that the prolonged X-ray emission would impose on the atmosphere. In their review of astrophysical ionizing radiation and Earth, \citet{Melott2011} establish the threshold for an “extinction-level event” as a globally averaged ozone depletion of about $30\%$. \citet{ThomMel2005b} argue that this would “nearly double the mean UVB flux at the surface” and could potentially trigger a food chain crash in the oceans.

So long as the ionizing radiation carries sufficient energy to break the strong N$_2$ bond and the generated NO$_x$ compounds reach an altitude equal to ozone abundances, it can initiate the catalytic cycle of ozone depletion. Thus, for our assessment of the threat imposed by SNe X-ray outputs, we follow a similar quantitative approach that these studies have used to characterize such damage, adopting their established parameterizations for atmospheric conditions on Earth and the necessary energy thresholds to trigger ozone depletion. Notably, however, the efficiency and altitude at which the ionizing radiation penetrates into the atmosphere will be dependent on the photon energy. The characteristic energy spectrum of an X-ray-luminous SN is substantially softer than that of a GRB and certainly lower than that of the high-energy particles associated with the later cosmic rays. We therefore must take into account modern-day Earth’s relative opacity to X-rays to properly generalize the ozone-related effects.

\subsection{Fluence and Lethal Distance}

In general, the amount of ozone depletion induced by an astrophysical ionizing event is mainly dependent on the spectrum and total amount of radiation incident on Earth — not the rate or duration of the event \citep{Ejzak2007,Melott2011} — and will be a function of the fluence, $\flu$, which is the energy deposited per unit area of the atmosphere. This is simply the integrated flux of radiation arriving at the top of the planet’s atmosphere. An explosion at distance $d$ with an X-ray flux $F_{\rm X}$ has an X-ray fluence, $\flu_{\rm X}$, of
\begin{equation}
\label{eq:fluence}
    \flu_{\rm X} = \int_{t_i}^{t_f} F_{\rm X}(t) \ dt = \frac{\int_{t_i}^{t_f} L_{\rm X}(t) \ dt}{4\pi d^2} = \frac{E_{\rm X}}{4\pi d^2} \ \ .
\end{equation}
We see that the fluence depends on the total X-ray energy output in the observed window: 
\begin{equation}
\label{eq:EX}
    E_{\rm X} = \int_{t_i}^{t_f} L_{\rm X} \  dt
\end{equation}
which is the X-ray luminosity integrated over time.

In principle, the X-ray fluence in Equation~(\ref{eq:fluence}) could be diminished by absorbing material along the sight line to the SN. However, for the $\la 50 \ \rm pc$ distances of interest, this is likely a negligible effect, and we can thus appropriately adopt the unabsorbed luminosity when given.

To compare the magnitude of the threat associated with the X-ray emission of each SN, we will calculate the furthest distances at which they could be from an Earth-like planet to impose lethal effects on the biosphere. A simple comparison point is the “lethal distance,” which is the terminology used throughout \citet{ThomMel2005a,ThomMel2005b} and \citet{Melott2011,melthom2017} and which we define as the approximate distance at which an SN would impose severe lethality on a terrestrial biosphere. We will thus characterize the degree of damage by a critical fluence, $\flucrit_{\rm X}$, in the X-ray band.  

Demanding that the fluence in Equation~(\ref{eq:fluence}) be equal to the critical fluence at which the X-ray emission would impose lethal consequences, we solve to find the associated distance as
\begin{eqnarray}
\label{eq:lethdist}
    \lethdist_{\rm X} & = & \left( \frac{E_{\rm X}}{4\pi \flucrit_{\rm X}} \right)^{1/2} \\
    & = & 14 \ {\rm pc} \ 
    \pfrac{E_{\rm X}}{10^{49} \ \rm erg}^{1/2} \ 
    \pfrac{400 \ \rm kJ \ m^{-2}}{\flucrit_{\rm X}}^{1/2}
\end{eqnarray}
To then evaluate the range of influence for X-ray SNe, we must (1) appropriately parameterize the critical fluence value for the soft X-rays of the SNe data and (2) find each SN’s total observed X-ray energy output, $E_{\rm X}$. The remainder of this section is devoted to the discussion of critical fluence and its appropriate value for X-ray emission. Total SN X-ray energy outputs will be calculated and discussed in \S \ref{sect:data}. 

\subsection{Critical Fluence}

Table \ref{tab:SNproperties}, column 3, displays the observed energy band for the SN X-ray observations in each paper. We note that these observations are mainly limited to the soft X-ray energy band ($\leq$ 10 keV). Soft X-ray photons each carry significantly lower energy than the gamma rays and cosmic rays that most prior studies relevant to nearby SNe have focused on. Moreover, the duration of the X-ray radiation (months to years) is longer than that of a GRB but significantly shorter than that of the cosmic rays that arrive later (hundreds to thousands of years). These factors must be accounted for to appropriately assign a critical fluence value to the X-ray lethal distance.

Importantly, Earth’s thick atmosphere is more or less opaque to soft X-rays, meaning that the photons arriving at the top of the atmosphere will not reach the surface, and it is unlikely that the majority of these photons make it down to the stratospheric level of the ozone layer. Nonetheless, due to the sheer magnitude of energy associated with SNe, even the softer X-rays can make it to an altitude adequate to initiate the NO$_x$-induced catalytic cycle of ozone depletion — a process further dependent on meteorological conditions that facilitate the downward transport of NO$_x$ into the stratosphere \citep{Solo1982}. This is an important caveat and will be addressed in more detail in the next subsection.

Further complicating the matter, a number of planetary conditions can play a role in determining the impact of astrophysical ionizing radiation on the atmosphere. Time of year, angle of incidence, geomagnetic activity, and meteorological conditions are just a few of the many influences determining the Earth system’s response \citep{Tart2020}. Previously published work has addressed these factors through the use of atmospheric chemistry models that can calculate globally averaged ozone depletion.

The most useful of these studies with respect to ozone-related effects and astrophysical events are \citet{ThomMel2005a,ThomMel2005b} and \citet{Ejzak2007}. These papers each utilized an atmospheric model of modern-day Earth (the Goddard Space Flight Center two-dimensional atmospheric model) to explore the atmospheric effects associated with ionizing radiation from astrophysical sources. Since they both provide insight into the fluence ranges and sensitivity to photon energy, their results prove vital for our analysis.

\citet{ThomMel2005a, ThomMel2005b} studied the relation between ozone depletion and critical fluence. They compared the response to ionizing radiation with three different fluence values (10 kJ m$^{-2}$, 100 kJ m$^{-2}$, 1000 kJ m$^{-2}$), finding that the percent change in the globally averaged column density of ozone scales with fluence as a power law with an index $\sim 0.3$. This less than linear relationship occurs due to the ozone depletion becoming saturated at higher fluence values. While this work specifically used timescales associated with GRBs, its general implications for ionizing radiation remain the same, and it provides necessary context for scaling our critical fluence value.

\citet{Ejzak2007} then utilized the same atmospheric model to examine the terrestrial consequences of spectral and temporal variability for ionizing photon events. In this study, they varied the burst duration (10$^{-1}$--10$^8$ s) and average photon energies (1.875 keV – 187.5 MeV), while holding fluence constant at 100 kJ m$^{-2}$. They found that at a given fluence, higher-energy photons were more damaging to ozone because they penetrate deeper into the atmosphere, creating a significant increase in NO$_x$ at stratospheric altitudes. 

Nonetheless, at the fluence used in the study, gamma rays with energies 187.5 keV and above destroy $\gtrsim 33\%$ of the ozone. This is sufficient to represent a lethal dose, particularly for the most vulnerable biota. For this reason, $\flucrit_{\gamma} = 100 \ \rm kJ m^{-2}$ is often assigned to be the critical fluence for gamma-ray exposure \citep{Melott2011}.    
Turning to X-rays, \citet{Ejzak2007} found that a photon energy of 1.875 keV (which most closely corresponds to the soft X-ray bands cited here) induces a globally averaged ozone depletion of $\sim 22\%$ (at the constant 100 kJ m$^{-2}$ fluence). This depletion level would persist for a couple of years, begin to noticeably recover after about five years, and effectively complete recovery in a bit over a decade — well before the later arrival of the cosmic rays. While this may seem relatively short on geological timescales, it is nevertheless many generations for the UV-transparent single-celled organisms at the base of the marine food chain and could potentially leave measurable traces in the paleontological record \citep{cockell99}. For reference, the peak of anthropogenic-related ozone destruction was around $\sim 5\%$ globally averaged depletion in the 1990s \citep{WMOrep}, so we are discussing average depletion levels significantly higher than any recent phenomena.

Still, the globally averaged ozone depletion of $\sim 22\%$ at an X-ray fluence of $100 \ \rm kJ m^{-2}$ would be very unlikely to induce severe lethality or serve as a lone trigger for an extinction-level event and thus would not fit the parameterization for a lethal distance. Therefore, a fluence value of $\sim$ 100 kJ m$^{-2}$ is too small for the softer X-ray band. A higher critical fluence value is instead required for the given photon energies and burst durations of X-ray-luminous SNe. To account for this, we combine the modeled results from both \citet{Ejzak2007} and \citet{ThomMel2005a,ThomMel2005b} (ozone depletion and fluence scales less than linearly) and adopt a higher critical fluence value of 
\begin{equation}
\label{eq:FlucritX}
\flucrit_{\rm X} \approx 400 \ \rm kJ \ m^{-2} = 3.8 \times 10^{45} \ \rm erg \ pc^{-2}
\end{equation}
to induce $\gtrsim 30\%$ prolonged ozone depletion. With this value, an X-ray-luminous SN located at the typical SN lethal distance cited in \citet{Melott2011} ($D =$ 10 pc) would require a total X-ray energy output of $E_{\rm X} = 4.8 \times 10^{48}$ erg.

\subsection{Limitations and Caveats to Critical Fluence Calculation}

The percentages estimating biological damage are not precise, as a variety of conditions within the Earth system could alter the exact effects of ionizing radiation. The upshot of this is that the critical fluence value of 400 kJ m$^{-2}$ is also somewhat of a crude approximation, and the definition of a “lethal distance” is somewhat loose. However, these definitions and values remain sufficient to provide us with appropriate approximations for our threat assessment conducted in this paper, the purpose being to identify whether X-rays from interacting SNe exhibit a notable range of influence. A more extensive analysis of the associated atmospheric effects and variations in biological damage can be found throughout the aforementioned literature.

To put the fluence value of 400 kJ m$^{-2}$ into the context of an X-ray influx that has impacted Earth before, we can reference the largest solar flare on record, the 1859 Carrington Event. This event is typically classified as an X-45 solar flare for its peak flux of $\sim$45 $\times$ 10$^{-4}$ W m$^{-2}$ in soft X-rays \citep{Clive2013}. Following the calculations from Equations~(\ref{eq:fluence}) and (\ref{eq:EX}), this peak emission would have to persist from the Sun at that magnitude for approximately 2.8 yr to induce a fluence of 400 kJ m$^{-2}$. Moreover, this peak is in the soft X-ray band, without substantial evidence of significant emission in the 10+ keV range that characterizes interacting SN emission (see \S \ref{sect:data}). Nonetheless, solar flares can provide us with concrete empirical evidence of influxes of soft X-ray emission perturbing Earth’s atmosphere.

Further limitations in our approximation for the critical fluence and associated effects stem from the fact that no agreed-upon spectral model for X-ray-luminous SNe exists, and thus, we are extrapolating from modeled effects of events such as GRBs, which would have notably harder spectra than would be characteristic of the X-ray phase of emission from an SN.

The most notable discrepancy in using these spectral models would be with respect to the altitude of energy deposition in the atmosphere. In short, this complicates any direct extrapolation of the \cite{Ejzak2007} 1.875 keV results for our assessment. \citet{Ejzak2007} used a GRB spectrum, which they modeled as a broken power law in photon energy, with an adjustable peak energy in which the spectral index changes. A consequence of this power-law spectrum is that there are always photons with energy $> 10$ keV. A corollary of this is that when \citet{Ejzak2007} calculate the ozone depletion at a peak photon energy of 1.875 keV, there is still a significant proportion of higher photon energies (10+ keV) being inputted as well. This peripheral energy falls outside the spectral range of the confirmed SN observations. However, a good portion of the energy is within the range of all of the X-ray data we use. Moreover, if only considering the $<$ 10 keV photons, this would be a quite conservative approach, because it is very likely that substantial X-ray emission exists for each of these SNe outside their cited energy ranges, i.e., in the harder X-ray band that most X-ray telescopes cannot detect ($>$ 10 keV).
NuSTAR, the newest X-ray telescope that actually has the capability of sensing the more energetic photons ($\sim$ 3--79 keV), has already observed significant X-ray emission in these harder spectra for some SNe \citep{2015ApJ...810...32C, Thomas22,Bret2022}, and there is nothing to indicate this is abnormal. \citet{Ejzak2007} confirmed that harder spectra would result in more ozone depletion and would thus require a lower critical fluence to induce lethal effects on the biosphere. We therefore aimed to ensure that our approximation is conservative so as to not overestimate the SN influence.

\begin{figure}
    \centering
    \includegraphics[width=0.8\textwidth]{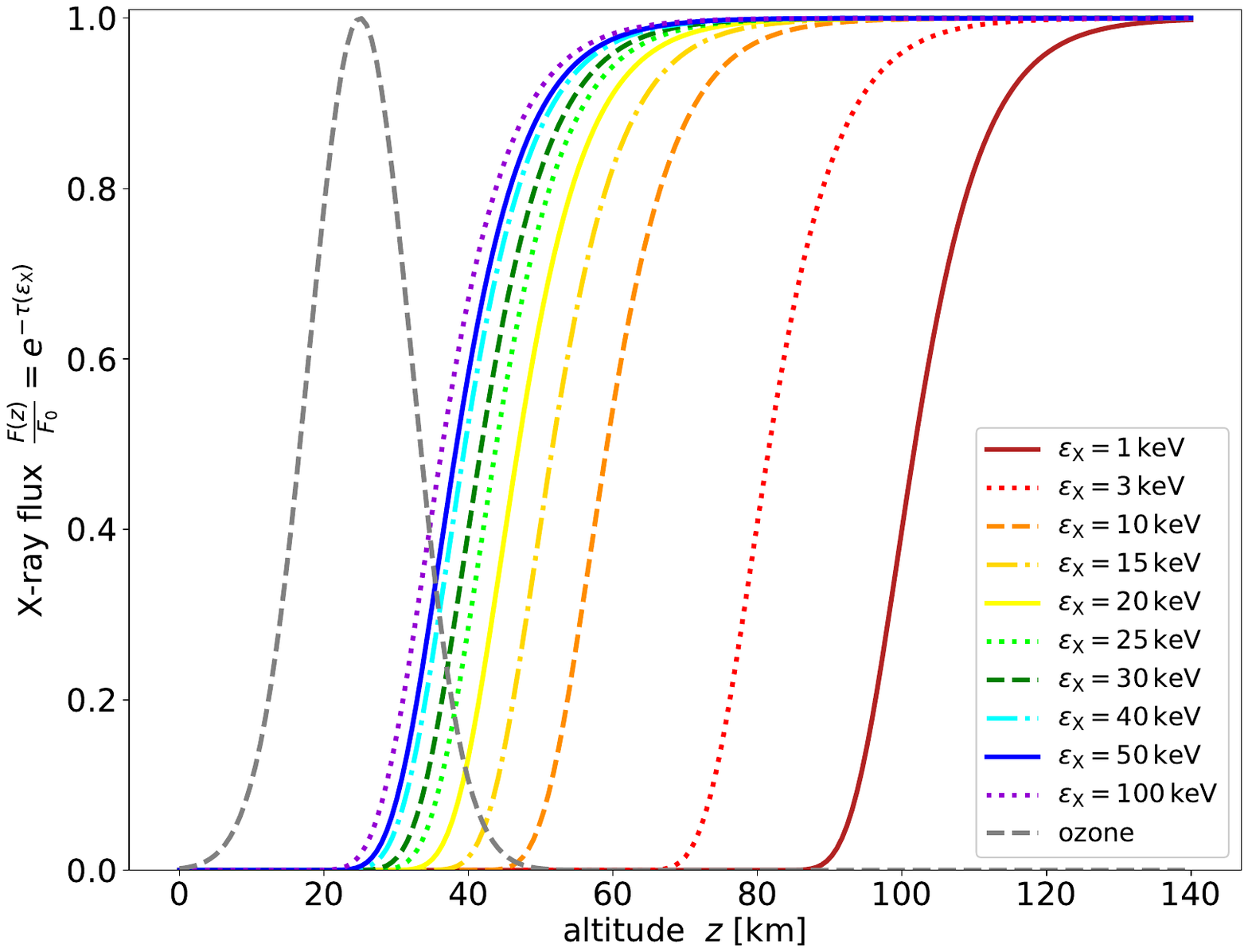}
    \caption{Profiles of X-ray flux attenuation, in an exponential model of Earth's atmosphere. Solid curves show the ratio $F(z)/F_0$ of X-ray flux $F(z)$ at altitude $z$ relative to the incident flux $F_0$, for photons with indicated energies.  The profiles start at unity for large $z$, then drop due to atmospheric absorption, which depends sensitively on X-ray energy. The gray dashed Gaussian curve is an idealized sketch of the average (unperturbed) stratospheric ozone profile.  We see that soft X rays with $\varepsilon_X \lesssim 10 \ \rm keV$ are stopped at altitudes far above the ozone layer, while higher-energy photons have progressively more overlap with the ozone.  We therefore must account for the reduced destructive efficiency of lower-energy X-rays, as in Equation~(\ref{eq:FlucritX}).
    }
    \label{fig:XrayFluxProf}
\end{figure}

We turn now to the atmospheric deposition of X-rays and the resulting ozone damage. X-rays are less penetrating than gamma rays generally, and X-ray absorption is a strong function of photon energy, with opacities dropping as $\kappa_X \propto \varepsilon_{\rm X}^{-3}$. Thus, low-energy (soft) X-rays are absorbed high in the atmosphere, while harder X-rays penetrate more deeply. We therefore expect soft X-rays to be less effective ozone depletion agents than harder X-rays and gamma rays.

Figure \ref{fig:XrayFluxProf} illustrates this behavior, showing atmospheric transmission for a range of X-ray energies $\varepsilon_{\rm X} \in [1,100] \ \rm keV$, compared to an indication of ozone layer. Plotted is the flux $F(z)$ at height $z$ relative to the incident flux $F_0$ at the top of the atmosphere. We have $F(z)/F_0 = e^{-\tau(\varepsilon_{\rm X})}$, with the optical depth $\tau(\varepsilon_{\rm X}) = \kappa(\varepsilon_{\rm X}) \ \Sigma(z)$. Here the opacity $\kappa$ is from the dry air tabulation of mass absorption coefficients by NIST database \citep{Hubbell2004}\footnote{\href{http://physics.nist.gov/xaamdi}{http://physics.nist.gov/xaamdi}}, and we use a simple exponential model for the atmosphere to evaluate the mass column density $\Sigma(z) = \int_z \rho(z^\prime) \, dz^\prime$.
We see that in all cases the flux ratio starts at unity at high altitudes then drops rapidly at a characteristic height that is energy dependent. In particular, the flux of $\varepsilon_{\rm X} = 1$ keV and 3 keV photons is cutoff below $\sim 70$ km, substantially above the majority of the ozone layer, roughly plotted by location in dashed gray. Photons with 10 keV and above overlap progressively more with the ozone.

The justification for considering ozone depletion, however, can be verified with empirical evidence of the photochemical coupling between the lower thermosphere and the upper stratosphere of Earth’s atmosphere \citep{Solo1982,Rand2006}. Essentially, most ionization of N$_2$ induced by the photons, as well as the subsequent creation of NO$_x$, will occur above the stratosphere — primarily in the lower thermosphere at altitudes above $\sim$90 km. In fact, it is the variability in the flux of the Sun’s soft X-rays that serve as the primary control for nitric oxide density in the lower thermosphere of the tropics \citep{Barth99}. However, \citet{Solo1982} first resolved that substantial amounts of NO$_x$ produced in the thermosphere can reach the stratosphere and trigger the catalytic cycle of ozone depletion. This descent of NO$_x$ from the thermosphere to the stratosphere has since been both modeled and consistently observed in a magnified response to solar flares \citep{Rohen2005,Rand2006,Mal2021,Bailey2022,Sisk2022}. The process is particularly prevalent at high latitudes during polar winter, with the vertical transport of NO$_x$ being further facilitated by various meteorological conditions, such as polar vortices. We therefore would expect that much of the global ozone depletion resulting specifically from SN X-rays would actually be contingent upon this photochemical coupling and would often be concentrated over the polar regions. The specifics of the depletion timeline and duration would be further dependent on the varying seasonality and geophysical conditions throughout the months and years of prolonged X-ray emission, with some months of the event being more/less lethal than others.

All considered, these variables are why the \citet{Ejzak2007} results are so useful, as they provide a general assessment, and their inputs closely align with the X-ray emission observed. One small caveat noted by the authors is that the direct relation between fluence and ozone depletion begins to weaken for burst durations longer than 10$^8$ s ($\approx$ 3 yr), as meteorological conditions (e.g. rainout) begin to remove NO$_x$ compounds from the atmosphere and thus dampen any prolonged increase in ozone depletion. Many of the SN X-ray emission profiles we analyzed have longer durations than 10$^8$ s, which would seem to indicate that more consideration is needed than simply finding their $E_{\rm X}$ value. However, for each SN analyzed, either all or the vast majority of their total energy output ($>$ 95$\%$) occurs well within this timeframe of 10$^8$ s, and therefore, effects such as rainout will not significantly alter our calculation of SN influence. In the future, an atmospheric model simulation that is specifically attuned to the associated energy inputs for SN X-ray light curves would be needed to refine speculation any further. For our threat assessment conducted here, this approximation suffices, and we adopt the critical flux in Equation~(\ref{eq:FlucritX}) as a rough indication of the effect of SN X-ray irradiation.

With an analytical method established for comparing X-ray emission effects, we now turn to the empirical data for SN X-ray emission collected in the last half-century.
\newpage

\section{Data: X-Ray-luminous Supernovae}
\label{sect:data}

The explosion processes governing SNe can be broken into two distinct origins: core-collapse SN (CCSN) and thermonuclear SN \citep{alsmurd2017}. CCSN occur at the end of a massive star’s ($\gtrsim 8 M_\odot$) lifetime, whereas thermonuclear SN typically occurs from white dwarfs accreting mass from a binary companion. Both processes have similar explosion energies overall, but notably, for our purposes, the magnitude and timing of X-ray emission will vary in relation to stellar mechanisms \citep{Smith2014,Burrows21}. 

Not all SNe show evidence for substantial outputs of X-ray emission. Instead, X-rays are primarily the consequence of the interaction between the expanding SN ejecta and the progenitor star’s circumstellar medium (CSM). The density of this CSM is directly related to the progenitor’s mass loss during the late stages of its stellar evolution. In general, a higher-density CSM will result in a greater magnitude of thermal X-ray emission \citep{Smith2014,Chandra18, Dwark2019}.

Stellar theory posits that the prerequisite condition for a dense enough CSM that leads to high X-ray emission is most likely — and perhaps only possible — in a CCSN event. Type IIn SNe, as we will see below, are the most X-ray-luminous SNe, and they show evidence for particularly strong circumstellar interactions due to high mass loss in the years and centuries leading up to the explosion \citep{GalYam2007,Smith2008,Kiewe2012}. The energy source and light-curve behavior of Type IIn events stem from the collision and shock of the blast with the CSM, and thus \citet{Smith2017} argues that they are best thought of as an “external phenomenon” rather than an SN type. Recent X-ray-bright events also include other SN classes, with multiple X-ray detections of Type Ib/c CCSNe \citep{Marg17,Chandra20,Thomas22,Bret2022}, as well as a Type Ia thermonuclear SN \citep{Boch18}.

The confirmed presence of relatively large X-ray emission across different SN classes is an exciting development in our overall understanding of the rates of stars’ mass loss and general SN physics. A detailed understanding of this prevalence, however, remains unclear. Whether the X-ray emission originates from a CCSN or a thermonuclear SN is not particularly relevant to the hazardous effects imposed on a terrestrial atmosphere that we are concerned with in this study. So long as the event manages to induce a high-enough fluence on the atmosphere, the associated impacts of the X-ray photons will be the same as a function of $E_{\rm X}$ and can be similarly calculated with Equation~(\ref{eq:lethdist}). As such, we will not be providing further analysis or speculation as to how and why these X-ray emissions occur, as more extensive discussions can be found within the above-cited literature. 

Note also that it is quite plausible that X-ray emission from SNe is not isotropic because the circumstellar medium may well be anisotropic. This means that the blast interaction could lead to different X-ray luminosities in different directions. In the absence of a detailed model of the circumstellar medium, there is no clear way to account for this for individual events. However, if there is anisotropy, this will be encoded in the frequency of observed X-ray-bright SNe, which we use in estimating the rates and threats posed by these events. This also means that these effects will be better accounted for with a larger and more complete observational survey of X-ray SNe.

All considered, the rate of occurrence for X-ray-bright SNe would have general implications for the Galactic habitable zone, i.e., the locations in which life could exist \citep{Line2004,Gowan2011,Cockell2016}. Therefore, when assessing the risks imposed on specific terrestrial biospheres, we need not discriminate between the type of X-ray luminous SN and its progenitor. But, to generalize our results and gain insight into planetary habitability, the different magnitudes at which each specific type of SN emits in the X-ray band are important, as this would factor into the rate of occurrence for dangerous X-ray emission throughout the universe (see \S \ref{sect:res}). 

\subsection{X-ray Data Analysis}

\begin{figure}
    \centering    \includegraphics[width=1\textwidth]{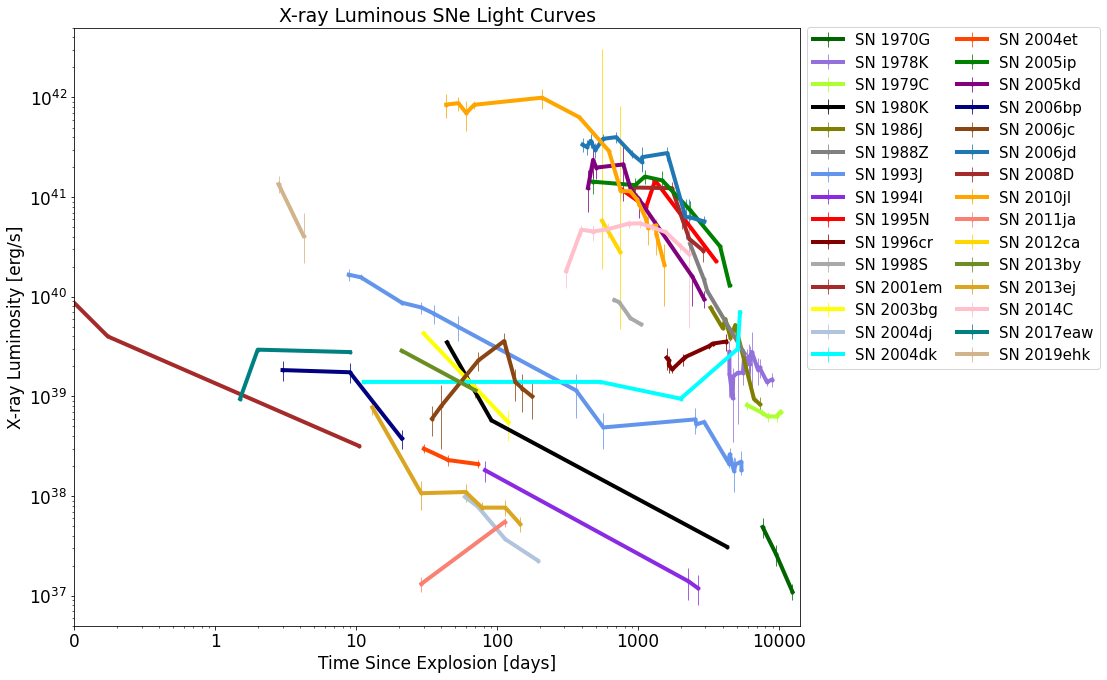}
    \caption{Displays the X-ray light curves for 30 of the 31 SNe analyzed (SN 1987A has been omitted for visual clarity due to its low luminosity). All data points are sourced from the papers listed in Table \ref{tab:SNproperties}. Both axes are logged; error bars are plotted when provided by the original paper. The figure further illuminates the range of available observations we have for each SN. Our data and subsequent analysis are limited to only these confirmed observational data points of X-ray emission.}
    \label{fig:SNe Light Curves}
\end{figure}

Figure \ref{fig:SNe Light Curves} displays the X-ray light curves for 30 of the 31 SNe that we identified in Table \ref{tab:SNproperties} (SN 1987A has been removed for clarity; see caption). These curves illustrate the evolution of each SN’s X-ray emission over time in the cited observed band, with each point representing a confirmed luminosity value taken from the sources cited in Table \ref{tab:SNproperties}. Both axes are logged. The figure further illuminates the often-limited range of available observations for each SN in the epoch they are observed.

To our knowledge, the most comprehensive depictions of such SN X-ray emission prior to this study were the SN X-ray database\footnote{\href{https://kronos.uchicago.edu/snax/}{https://kronos.uchicago.edu/snax/}} \citep{Ross2017, Nisenoff2020} and the \citet{Dwark12} X-ray SNe light curve compilation (and extended in \citet{Dwark2019}). We utilized these resources as an initial guide for our data collection and analysis, but then scoured the original papers and observational reports (cited in Table \ref{tab:SNproperties}) to compile our luminosity data and expand the scope of the analysis. 

Often the original papers directly provided luminosity calculations for the observed X-ray emission, from which we would directly adopt their derived values. When only an X-ray flux was given, we conducted the simple luminosity calculation using the adopted distance cited in the paper ($L_{\rm X} = 4\pi d^2 F_{\rm X}$), ignoring any effects of absorption due to the relatively short (astronomically) distances we are concerned with here.

From these data, we can calculate each SN’s total X-ray energy emitted, $E_{\rm X}$,
through an integration of their respective X-ray light curves as seen in Equation~(\ref{eq:EX}).
To do this, we take a simple empirical approach: we interpolate linearly between the observed epochs. We have tried other methods including using simple fitting functions and find that these give similar results. As described in \S \ref{sect:meth} (see Equation~(\ref{eq:lethdist})), we are using this total energy output to characterize the general threat imposed by each SN. 

\begin{figure}
    \centering
    \includegraphics[width=0.8\textwidth]{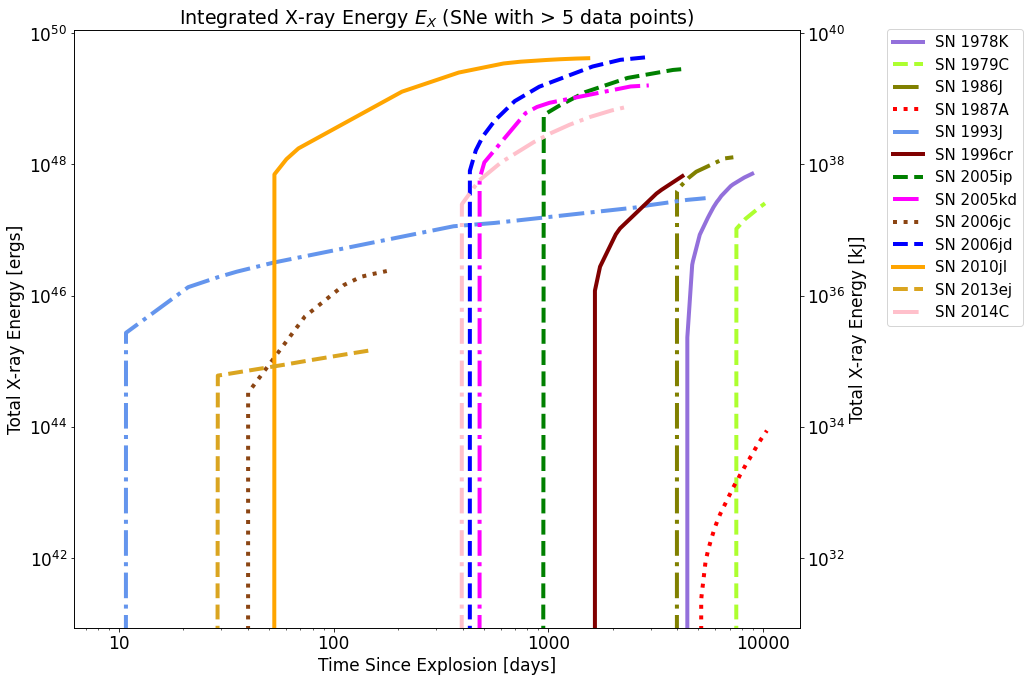}
    \caption{Integrated X-ray energy of SNe with more than five available data points (SN 2008D excepted). All data points are sourced from the papers listed in Table \ref{tab:SNproperties}. The observed X-ray luminosity is integrated via Equation~(\ref{eq:EX}) with linear interpolation between measured points. For each curve, the highest and latest point gives a lower limit to the total X-ray energy output $E_X$.}
    \label{fig:Integrated Energy Plot}
\end{figure}

Figure \ref{fig:Integrated Energy Plot} illustrates how these cumulative X-ray energy outputs evolve over time. For clarity, we limit the displayed SNe to those containing $> 5$ confirmed observations (with SN 2008D omitted for clarity) and only use measured data points that result in abrupt onset. We linearly interpolate the luminosity between the points, leading to the rise of the curves, which terminate at the final observation. The endpoint of each curve gives the total observed $E_{\rm X}$. The shape of the rise from onset to the final point indicates the time history of the radiation dose delivery.

The range of widths of the curves in Fig.~\ref{fig:Integrated Energy Plot} gives a general sense of the range of timescales associated with the measured SNe. The curves generally rise rapidly and then taper off, corresponding to a high initial X-ray luminosity that diminishes over time. In some examples, we have SNe measured at both early and late times (SN 1993J), which provide us with the understanding that these emissions are present for long periods of time after they first appear. Note that the time axis is logarithmic so the durations are not uniform widths across the plot.

A primary motive of our study is to work explicitly with confirmed SN data. Accordingly, two notable parameters are confining our analysis of the SN data to provide restrictive limitations in our overall assessment: (1) We are only considering the emission of these SNe from the given data points; i.e., we do not plot emission that may occur outside the epochs observed, and (2) we are only considering the emission of these SNe within the cited X-ray band reported by the original papers.

The likelihood that X-ray activity occurred outside the observational windows and outside the typical range of X-ray telescopic sensitivity ($\leq$ 10 keV) is very high, particularly for SNe with scant observational data. For example, in researching the surprising X-ray emission of Type Ib/c SNe, \citet{Marg17} have proposed that as much as $\sim$ 40\% of Ib/C SNe could be X-ray-luminous at t $\gtrsim$ 1000 days. \citet{Bret2022} reported on seven years of Chandra–NuSTAR observations of SN 2014C, which they note is a late interacting SN best modeled and observed at thermal emission with $T \approx 20$ keV. The implications of this continually growing evidence are that a significant percentage of SN X-ray emission is simply being unobserved, particularly in the shorter wavelengths. However, no agreed-upon spectral model for SN X-ray emission exists within the literature, and we make no attempt to adopt one here.

Therefore, with the limitations of (1) only observed epochs and (2) only observed energy ranges, all considered, our calculations will thus yield conservative values for the total X-ray energy outputs. We accept this given that here we are merely analyzing the empirical evidence to determine if these emissions can induce a notable impact with the confirmed energy alone and will extrapolate further in a future study. As the astronomical community gains further insight into the harder spectrum of X-ray emission and the X-ray observational cadence increases, we will be able to expand our assessment.

\subsection{Key Data Trends: SN Type and Emission Timescale}

\begin{figure}
    \centering
    \includegraphics[width=0.8\textwidth]{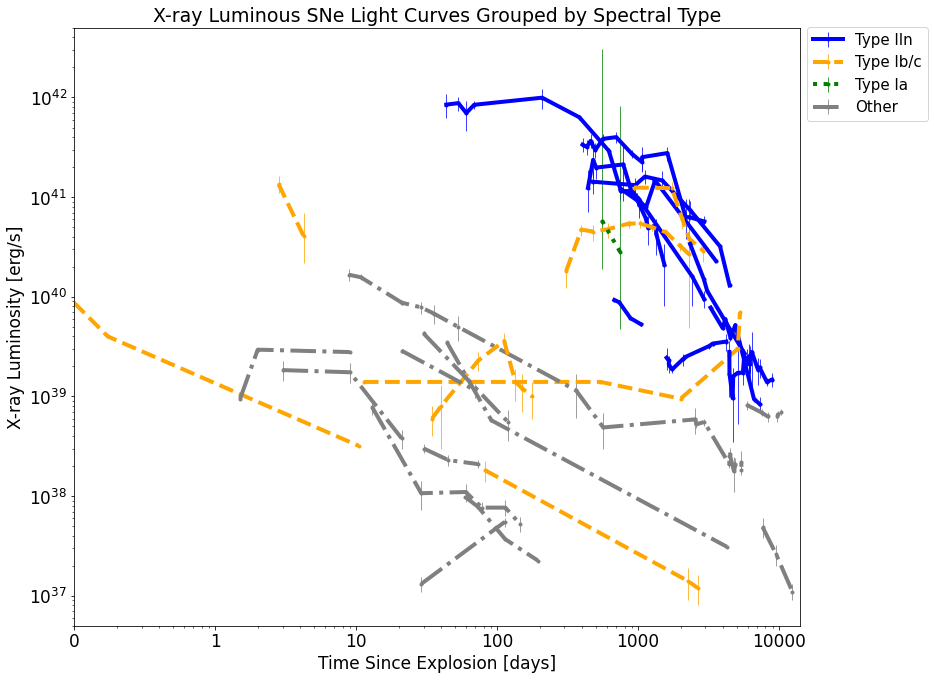}
    \caption{The same as Figure \ref{fig:SNe Light Curves}, with the light-curve coloring now corresponding to the spectral classification of the SN cited by the given paper(s). This highlights the X-ray emission disparities of the different spectral classifications, as well as the new evidence for X-ray emission among a variety of non-IIn types. All data points are sourced from the papers listed in Table \ref{tab:SNproperties}.}
    \label{fig:SNe Light Curves by Type}
\end{figure}

Figure \ref{fig:SNe Light Curves by Type} is the same as Figure \ref{fig:SNe Light Curves}, but with the light-curve coloring now grouped by spectral classification. This serves to highlight the X-ray emission disparities of the different spectral types. There are three key aspects emphasized here that are relevant to our threat assessment:

\begin{enumerate}
\setlength{\itemsep}{0pt}
\item 
The X-ray emission spectra from Type IIn SNe are significantly higher on average than all other types of SNe. Their X-ray emission is most readily seen months and/or years after the explosion. This is a characteristic of a progenitor star with a high rate of mass loss creating a dense CSM, with which the SN shock wave collides \citep{Dwark2019}. The rather late interaction reflects the travel time for the SN blast to reach the CSM. \citet{Chandra20}. Note however that this trend is partially attributable to observational biases, as some SNe IIn were not discovered or observed until well after their explosion dates (e.g., SN 1986J, 1988Z, and 1978K). Note also that most X-ray SN observations are sensitive to soft photons. The only Type IIn that has been observed by NuSTAR is SN 2010jl, which exhibited significant hard X-ray emission (10–80 keV) over two years after the explosion (we have not included this lone data point at that energy range since we are using only data sets with multiple confirmed observations). Regardless, the high, prolonged X-ray energetics imply that SNe IIn typically have the largest range of influence on terrestrial atmospheres during this phase of the SN.
\item 
There are thus far three observed SNe that have been initially classified as non-IIn SNe yet have X-ray emission spectra comparable to those of the IIn class. These are SN 2012ca (Type Ia), SN 2001em (Ib/c), and SN 2014C (Type Ib). The evolution of their light curves calls into question their initial classifications, as well as the traditional spectral classification system as a whole \citep{Marg17}. \citet{Chandra20} propose that these SNe have essentially “metamorphosed” into Type IIn SNe, transitioning from noninteracting to interacting SNe. Their high X-ray energetics potentially make them as lethal to nearby biospheres as the IIn class. Furthermore, SN 2014C was observed for several years with NuSTAR (range 3–80 keV) and showed significant luminosity output in the hard X-ray spectrum throughout \citep{Thomas22,Bret2022}. The evidence for these harder emissions elsewhere would have implications for the critical fluence value discussed in \S \ref{sect:meth}, as the amount of ozone depletion scales with photon energy and thus would ultimately serve to increase their range of influence. Whether the characteristics of these non-IIn spectra are typical for those of their class could also have significant consequences for the habitable zone, as a higher rate of occurrence for X-ray emission could greatly increase the statistical threat that SNe pose for planetary habitability (see \S \ref{sect:res}).
\item
The X-rays originating from other types of SNe (IIL, IIP, etc.) are relatively small in comparison. These SNe are “noninteracting” in the absence of a dense CSM, and the X-ray luminosities peak upon shock breakout at the considerably lower durations required to make their X-ray emission a threat to terrestrial biospheres. Though the luminosities in this shock breakout can be relatively large, they do not persist long enough to be of any significance regarding the lethal effects evaluated here but could nevertheless impose ionospheric disturbances.
\end{enumerate}
This last point is best encapsulated by the observed data from SN 2008D (whose entire observed light curve is only partially displayed for clarity in Figures {\ref{fig:SNe Light Curves}} and \ref{fig:SNe Light Curves by Type}). SN 2008D is quite a unique event in X-ray astronomy, in that it was a serendipitous discovery of an SN upon initial outburst \citep{Sod2008}. This offered a view of the SN breakout immediately in the X-ray band. Its peak observed X-ray luminosity of $\sim 3.8 \times 10^{43}\text{ erg s}^{-1}$ \citep{mod2009} is actually the most luminous observation in our data — nearly two orders of magnitude larger than the most energetic observations of Type IIn SNe. However, this burst of energy immediately dissipates within a matter of minutes by a few orders of magnitude, resulting in a relatively low total X-ray energy.

This bolsters a key notion underlying our analysis: For non-GRB SNe, the initial influx of photons from the SN outburst is nonthreatening to terrestrial biospheres at formidable distances. Instead, the prolonged X-ray emission that arises from CSM interaction (months/years after outburst) provides an additional threat and alters the timeline by which a nearby SN influences a terrestrial biosphere. In general, noninteracting SNe would not have much further lethal influence beyond the hazardous effects associated with their cosmic rays discussed in previous nearby SN studies.

\section{Results: Threat Assessment}
\label{sect:res}

Having characterized the effects of a lethal X-ray dose on the atmosphere and the output of the brightest X-ray SNe, we are now in a position to assess the threat that these explosions pose to the biosphere. Table \ref{tab:SNCalcs} displays the total X-ray energy output, $E_{\rm X}$, for the 31 SNe analyzed. Plugging these numbers into Equation~(\ref{eq:lethdist}), using the critical fluence value of 400 kJ m$^{-2}$ from Equation~(\ref{eq:FlucritX}), reveals the lethal distance, $\lethdist_{\rm X}$, for each SN. Over half of the SNe have lethal distances well over 1 pc. The 10 Type IIn SNe show the highest average range of influence, with lethal distances in the tens of parsecs. As derived from the likely ozone depletion that would occur at this fluence, these SNe are shown to pose a substantial threat to nearby habitable planets. This much is evident even with the restrictive limitations we have imposed in our analysis of the total X-ray emission (only confirmed observational epochs and energy ranges).

\begin{table}
\caption{Displays the Total X-Ray Energy Output, $E_{\rm X}$, for the 31 SNe We Have Analyzed, with the Corresponding Lethal Distance, $\lethdist_{\rm X}$, Found Using Equation~(\ref{eq:lethdist})}
\begin{center}
\begin{tabular}{|c c c c c c|} 
\hline
 Supernova & Classification & Integration Time & Total X-ray Energy & Lethal Distance & \\
 (Name) & (Spectral Type) & $t_f-t_i$ [days] & $E_{\rm X}$ [$\times10^{46} $erg] & $\lethdist_{\rm X} (\cal F)$ [pc] &\\
 \hline
 2006jd & IIn & 2537 & 4300 & 30 &\\
 2010jl & IIn & 1492 &	4200 &	29 &\\
 2005ip & IIn & 3978 &	2900 &	25 &\\
 1995N & IIn & 2795 &	2200 &	21 &\\
 2005kd & IIn & 2500 &	1600 &	18 &\\
 2001em & Ib/c & 1966 & 1400 &	17 &\\
 2014C & Ib & 1999 & 750 &	13 &\\
 1988Z & IIn & 3343 &	300 &	7.9 &\\
 1986J & IIn & 4022 &	130 &	5.2 &\\
 2004dk & Ib & 5266 &	82 &	4.1 &\\
 1978K & IIn & 4477 &	72 &	3.9 &\\
 2012ca & Ia & 191 &	71 &	3.9 &\\
 1996cr & IIn & 2625 &	67 &	3.7 &\\
 1993J & IIb & 5399 &	31 &	2.5 &\\
 1979C & IIL & 4310 &	26 &	2.3 &\\
 1998S & IIn & 370 &	22 &	2.1 &\\
 1980K & IIL & 4207 &	12 &	1.6 &\\
 2006jc & Ib & 143 & 2.4 & 0.71 &\\
 1994I & Ic & 2557 & 1.9 & 0.63 &\\
 2003bg & IIb & 90 & 1.9 & 0.63 &\\
 2019ehk & Ib & 1 & 1.1 & 0.47 &\\
 1970G & IIL & 4784 & 1.1 & 0.47 &\\
 2008D & Ib/c & 10 & 0.88 & 0.43 &\\
 2013by & IIL & 49 & 0.85 & 0.42 &\\
 2006bp & IIp & 18 & 0.20 & 0.21 &\\
 2017eaw & IIP & 8 & 0.18 & 0.20 &\\
 2013ej & IIP/L & 132 & 0.15 & 0.17 &\\
 2004et & IIP & 42 & 0.086 & 0.13 &\\
 2004dj & IIP & 135 & 0.052 & 0.10 &\\
 2011ja & IIP & 84 & 0.025 & 0.072 &\\
 1987a & IIP & 5397 & 0.0088 & 0.043 &\\
 \hline
\end{tabular}
\end{center}
{\bf Notes.} This distance is calculated as a function of the critical fluence value in Equation~(\ref{eq:FlucritX}), a conservative estimate for the soft X-ray energy spectrum (see \S \ref{sect:meth}). Importantly, the integration times, $t_f-t_i$, are the range of the observation, i.e., the time between the first and last confirmed data points. These are the times used to calculate $E_{\rm X}$ via Equation(\ref{eq:EX}). They are not necessarily an accurate representation of time since explosion or the total time of X-ray emission, as some SNe were not initially observed in the X-ray band until long after their initial outburst. Additionally, some SNe have wide gaps in observation in which the exact emission profile would go undetected.
\label{tab:SNCalcs}
\end{table}

\subsection{Ranges of Lethal Influence}

\begin{figure}
    \centering
    \includegraphics[width=0.8\textwidth]{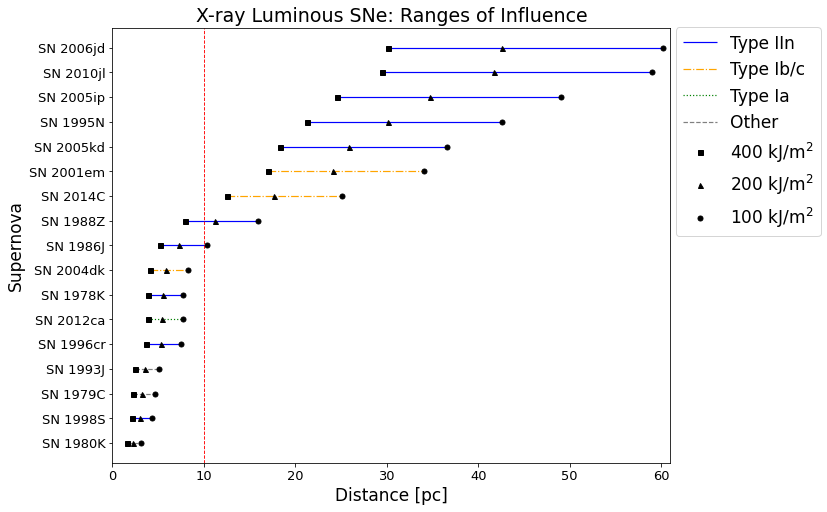}
    \caption{Displays the ranges of influence for all observed SNe with lethal distances of at least 1 pc. From left to right, symbols correlate to variation in critical fluence value used in Equation~(\ref{eq:lethdist}): 400 kJ m$^{-2}$ (square), 200 kJ m$^{-2}$ (triangle), and 100 kJ m$^{-2}$ (circle). The dotted red line denotes 10 pc, the previously cited average lethal distance for SNe from \citet{Melott2011}.}
    \label{fig:SN Range of Influence}
\end{figure}

Figure \ref{fig:SN Range of Influence} illustrates the approximate ranges of influence for the top 17 SNe analyzed (those with $\lethdist_{\rm X} > 1$pc). Here, we display three distinct values for a range of lethal influence, with each value derived from Equation~(\ref{eq:lethdist}), using three separate critical X-ray fluences (displayed on plot, left to right): $\flucrit_X = 400 \ {\rm kJ m}^{-2}$ (square), 200 kJ m$^{-2}$ (triangle), and 100 kJ m$^{-2}$ (circle). This serves to illustrate the distances at which these SNe would maintain significant influence on an Earth-like biosphere. 

We have gone to great lengths to justify that the leftmost value (400 kJ m$^{-2}$), which comes from Equation~(\ref{eq:FlucritX}), is most appropriate to calculate the lethal distance if the entirety of the X-ray emission was captured by confirmed observations. And therefore, as discussed extensively in \S \ref{sect:meth}, the square on each SN’s respective line corresponds to the range at which it would likely induce 30+$\%$ ozone depletion from merely the observed X-ray emission alone. 

But we have also shown throughout our discussion that it is very unlikely that the total X-ray emission from these SNe is encapsulated by the limited epochs and energy ranges thus far reported. So here, we now provide some appropriate — but still conservative — estimations for the potential ranges at which interacting SNe could be lethal.

The rightmost value displayed along the lines for each SN in Figure \ref{fig:SN Range of Influence} corresponds to a fluence of 100 kJ m$^{-2}$, which is often used in previous studies that utilize atmospheric chemistry models to analyze radiative effects on Earth-like atmospheres \citep{ThomMel2005a,ThomMel2005b, Ejzak2007}. In particular, \citet{Ejzak2007} specifically used this value with an average photon energy of 1.875 keV and a burst duration of $\sim$10$^8$ s, which are quantities that align most accurately with our given X-ray data. In these simulations, such an event induced globally averaged ozone depletion rates of approximately 22$\%$, followed by recovery timescales of over a decade. So at the marked distances of the circle on each SN’s line, these modeled effects would likely occur at the distance from merely the observed X-ray emission alone. Depletion of such scale is unlikely to trigger any type of extinction-level event and would be unlikely to fit the parameter for our lethal distance. Nevertheless, even at these levels, the event would be a substantial forcing on a terrestrial planet.

Of particular note is that if all X-ray-luminous SNe do indeed emit in the harder X-ray band ($>$ 10 keV) undetectable by most X-ray telescopes and do indeed emit outside the limited windows of our observations, then it is entirely appropriate to assign a critical fluence of 200 kJ m$^{-2}$ or even 100 kJ m$^{-2}$ for the lethal distance calculation of Equation~(\ref{eq:lethdist}). Figure \ref{fig:SN Range of Influence} shows that this adjustment would significantly expand the distances from which these SNe could impose lethal effects. And importantly, it is very likely the case, given (1) the confirmed evidence of hard X-ray emission that is seen by NuSTAR in SN 2010jl \citep{2015ApJ...810...32C} and SN 2014C \citep{Thomas22,Bret2022}, and (2) the likelihood that all X-ray-luminous SNe emit X-rays beyond the limited observed epochs (see \S \ref{sect:data}).

We have therefore shown that the most X-ray-luminous SNe can yield lethal distances of
\begin{equation}
\label{eq:lethaldistX}
    \lethdist_{\rm X} \ \sim \ 20-50 \ \rm pc
\end{equation}
and we have remained conservative in estimating their X-ray output. This range is larger than the canonical 8–10 pc value, which has significant consequences, as we will now see.

\subsection{Rates of Lethal X-Ray Supernovae}

Having shown that X-ray-luminous SNe can affect habitability at formidable distances, we now infer the implications that this will have for the overall Galactic habitable zone. The variables of most importance here are the rates of occurrence for interacting SNe, along with the geometry and dispersion of the galaxy itself. The large distances we found above correspond to a larger volume of influence, making the stage of SN X-ray emission all the more significant.

We wish to compute a rate $\Gamma_i(r)$ of SNe of Type $i$ within a distance $r$
(Sovgut et al., in preparation).  
This depends on the global Milky Way rate 
$f_X {\cal R}_{\rm cc}$ of X-ray-luminous SNe, which we write
as a product of the CCSN rate
${\cal R}_{\rm cc} = dN_{\rm cc}/dt$
and the fraction $f_X$ of SNe that
are X-ray bright.
The distances of interest are smaller than the scale height, $h_{\rm cc}$, for core-collapse progenitors; i.e., we have $r \la h_{\rm cc} \sim 100 \ \rm pc$.
In this limit, to a good approximation, we have
\begin{equation}
    \Gamma(r) \ \approx \ f_X {\cal R}_{\rm cc} \rho_{\rm cc}(R_\odot,z_\odot) \frac{4\pi r^3}{3}
    = \frac{r^3}{3 R_{\rm cc}^2 h_{\rm cc}} \
    e^{-z_\odot/h_{\rm cc}} e^{-R_\odot/R_{\rm cc}} \
    f_X {\cal R}_{\rm cc}
\end{equation}
where the observer's position is that of the Sun, namely, $(R_\odot,z_\odot)$ in Galactocentric coordinates.
We have assumed all core-collapse events follow the same “double exponential” distribution with probability density $\rho_{\rm cc}(R,z) = e^{-|z|/h_{\rm cc}} e^{-R/R_{\rm cc}}/4\pi R_{\rm cc}^2 z_{\rm cc}$, normalized to $\int \rho \ dV = \int \rho \ R \ dR \ dz \ d\phi = 1$.

For numerical values, we follow \citet{Murphey2021}. We adopt a solar distance, $R_\odot=8.7 \ {\rm kpc}$,
and height, $z_\odot = 20 \ \rm pc$.
We assume CCSN-like within a thin disk with scale radius and height, $R_{\rm thin}, h_{\rm thin} = 2.9 \rm \ kpc$, $95 \rm \ pc$, 
and we adopt the present Galactic core-collapse rates of ${\cal R}_{\rm cc} = 3.2^{+7.3}_{-2.6} \ \rm events/century$.
With these parameters, the “lethal rate” for non-IIn CCSN is
\begin{equation}
\label{eq:CCzap}
    \Gamma_{\rm CC}(r_{\rm CC}) =  0.5^{+1.2}_{-0.4}  \ {\rm events/Gyr} \  
    \left( \frac{r_{\rm CC}}{10 \ \rm pc} \right)^3
\end{equation}
where we adopt the fiducial lethal distance $r_{\rm CC} = 10$ pc suggested recently by \citet{Melott2011}. This gives a mean recurrence time $\Gamma_{\rm CC}^{-1} = 2 \ \rm  Gyr$, which gives about a 50\% chance that a lethal event has occurred in the $\sim 1$ Gyr history of complex life on Earth. Note the sensitivity to the adopted distance: If we instead use $r_{\rm CC} = 20$ pc as suggested by \citet{Thomas2023},
the rate jumps to $\Gamma_{\rm CC} = 4.0^{+9.2}_{-3.6} \ \rm Gyr$, and 
the recurrence timescale is only $\Gamma_{\rm CC}^{-1} \simeq 250 \ \rm  Myr$.

Now we assume Type IIn SNe have a lethal distance of $r_{\rm IIn}= 30$ pc and are a fraction $f_{\rm IIn} = 0.07$ of all core-collapse events (\citealt{Li2011}; also see \citealt{Cold2023}). 
Then, the “lethal rate” from these events is 
\begin{equation}
        \Gamma_{\rm IIn}(r_{\rm IIn}) =
        f_{\rm IIn} \Gamma_{\rm CC}(r_{\rm IIn})
     = 1^{+2}_{-0.8}  \ {\rm events/Gyr} \  
     \pfrac{f}{0.07}
    \left( \frac{r_{\rm IIn}}{30 \ \rm pc} \right)^3
    \label{eq:IInzap}
\end{equation}
This is twice the global average rate in Equation~(\ref{eq:CCzap})! This illustrates that Type IIn events pose an outsized threat despite their relatively modest rates of occurrence. The local core-collapse rate for all lethal core-collapse events is $\Gamma_{\rm tot} = \Gamma_{\rm CC} + \Gamma_{\rm IIn} =(690 \ \rm Myr)^{-1}$, assuming the fiducial lethal distances in Equations~(\ref{eq:CCzap}) and (\ref{eq:IInzap}).

As discussed in \S \ref{sect:data}, there is continuing evidence for non-IIn SNe evolving to show characteristics of interactions that result in high X-ray luminosities at late times \citep{Marg17,Chandra20, Thomas22}. This sort of evolution may be common for Type Ib/c SNe but simply missed by X-ray telescopes due to a lack of late-time observations. \citet{Marg17} estimate that as high as $\sim$40\% of Ib/c may evolve to be interacting. If these continue to have X-ray luminosities comparable to IIn explosions, this will have profound effects on the rates of lethality discussed here, since Type Ib/c events comprise about 19\% of all SNe and about 25\% of all CCSNe. This could then roughly triple the rate shown in Equation~(\ref{eq:IInzap}).

We also note that the result in Equation~(\ref{eq:IInzap}) depends sensitively on the Type IIn fraction and especially on their typical X-ray lethal distance. In particular, we see that the relative threat is set by the ratio
$\Gamma_{\rm IIn}/\Gamma_{\rm CC} = f_{\rm IIn} (r_{\rm IIn}/r_{\rm CC})^3$.
Clearly, further observations — particularly in the X-ray — will be critical to firming up this estimate and determining the true impact of X-ray-luminous SNe for Galactic habitability.

\section{Discussion}
\label{sect:disc}

Perhaps the most interesting results are the distances at which the X-ray emission can impose lethal effects on an Earth-like biosphere. This larger range of influence has consequences for the Galactic habitable zone, such as the harmful implications for recently discovered exoplanets that would be susceptible to nearby SNe \citep{Ramos23}. Importantly, this also opens the discussion for SN X-ray emission having directly influenced Earth’s own biosphere. The sheer magnitude of X-ray emission has both lethal and nonlethal implications relevant to Earth’s past, which we will briefly discuss in the remainder of this section, saving a more detailed analysis of the nonlethal effects on the atmosphere for a future study. 

\subsection{Implications for Earth's Past}

The empirical evidence for near-Earth SNe in the geologically recent past comes most readily from detections of \fe60 in the geological record. The recorded abundances have allowed for estimates of the distance from Earth at which the SN(e) likely occurred. These estimates range from 20 to 150 pc \citep{FieldsEllis1999, FieldsEllisHochmuth2005, Fry_2015}, and candidate star clusters have been proposed at distances around 50 pc and 100 pc \citep{Benitez2002,Mamajek2007,Hyde2018}. 
Remarkably, the lower end of this distance range includes the X-ray lethality distance in Equation~(\ref{eq:lethaldistX}). It is uncertain whether the SN Plio event $\sim 3 \ \rm Myr$ ago was a strong X-ray emitter. But if this event or any one of the $\gtrsim 10$ other SNe needed to form the Local Bubble were X-ray luminous, there could have been significant consequences. It is thus quite possible that SN X-ray emission imposed lethal effects on Earth organisms or, at minimum, once altered the Earth’s atmospheric ozone levels. 

For further comparison, previous assessments of SN ozone damage found a lethal distance ranging from $\lethdist \sim 8 \rm pc$ \citep{Gehrels2003} to $\sim 10 \rm pc$ \citep{Melott2011}, indicated by the dashed line (red) in Figure \ref{fig:SN Range of Influence}. These do not include the effects of high-energy cosmic-ray muons, which could extend this range. In a recent re-evaluation that considered both ozone and muon effects, \citep{Thomas2023} revised the lethal distance up to 20 pc. We note that in these estimates, the ozone damage due to cosmic rays dominates here because the gamma-ray energy $E_\gamma \sim 2 \times 10^{47} \ \rm erg$ and associated critical fluence $\flucrit_\gamma = 100 \ {\rm kJ m^{-2}}$ give only $\lethdist_\gamma \approx 4 \ \rm pc$. We further note, however, that the gamma-ray emission mostly arises from the radioactive decays of $\iso{Ni}{56} \stackrel{\small 9 \ \rm days}{\longrightarrow} \iso{Co}{56} \stackrel{\small 111 \ \rm days}{\longrightarrow} \iso{Fe}{56}$ and thus is proportional to the \iso{Ni}{56} yield. SNe with large \iso{Ni}{56} production, such as Type Ia events, will have a larger gamma-ray fluence and thus present a greater threat during this phase.

We must also again draw attention to the fact that the later cosmic rays would linger for substantially longer than the duration of X-ray emission (see Figure \ref{fig:SN Evolution}). So while it seems readily apparent that the X-ray-luminous SNe would impose their lethal effects at larger distances from their X-ray emission alone, it remains an open question for further research as to how exactly the specifics of this lethality compare to that associated with cosmic rays alone.

\subsection{Other Influences on the Atmosphere}
\label{sect:other}

To this point, we have been primarily considering lethal effects related to ozone depletion, as this is a common point of comparison among the previous literature discussing the lethality of SNe. Importantly, however, there will be other effects that would pose interesting consequences for a terrestrial atmosphere or a planetary system in general. We reserve a more detailed assessment of these effects for a later study but can make general comments about the likely scenarios here.

Numerous studies have observed and modeled the middle atmosphere effects with respect to X-class solar flare events, which are characterized by an influx of solar soft X-ray emission \citep{Veron2002,pett2018,Sisk2022}. In addition to the already mentioned increase in odd nitrogen production in the mesosphere and lower thermosphere, the enhanced photoionization resulting solely from the soft X-ray enhancement of a flare can cause a sudden ionospheric disturbance, which would primarily deliver a “dramatic increase in electron density” and an “altitude-dependent temperature increase” \citep{pett2018,Mitra74,Hayes2017}. This ionization is sensitive to even small-scale changes in X-ray activity and would be most influential on the \emph{D} and \emph{E} regions of the ionosphere. The consequences of this on Earth are not necessarily relevant to biological lethality; however, the enhanced ionization would be measurable by modern instrumentation and likely have particular relevance for technologically advanced civilizations as these enhancements can have a considerable impact on radio communications, astronaut health, satellite degradation, and more.

A relevant point of comparison for these sudden ionospheric disturbances is with respect to solar flares. \citet{Clive2013} offered a detailed analysis of the largest solar flare on record, the 1859 Carrington Event. Here they used the data from \citet{Veron2002} to determine the total soft X-ray fluence of this event as 6.4 J m$^{-2}$ — the largest in recorded history that some speculate was matched by the large X-class solar flare of 2003 \citep{Curt2016}. An interesting thought arises here that contextualizes the shear magnitude of X-ray energy we have been considering in these SNe. Let us take the median of the SN IIn, SN 2005kd, which had a total observed X-ray energy of 1600 $\times$ 10$^{39}$ J (Table \ref{tab:SNCalcs}). If we now demand that the fluence in  Equation~(\ref{eq:lethdist}) match only that of the Carrington Event’s X-ray fluence, we derive a distance of 4600 pc. Now, utilizing the same parameters in Equation~(\ref{eq:IInzap}) and demanding the distance be the 4600 pc, we calculate a recurrence interval of $\sim$4 events per millennium. At these distances, however, dust extinction would now have to be taken into account — particularly for an SN within the galactic plane — serving to lessen the magnitude of X-rays arriving at the planet. Thus, further absorption effects would need to be considered, and in all, the comparison is not exact, but this nevertheless offers a simplified clarification for the distances at which the X-ray emission could perturb the atmosphere.

At this low of a fluence, no lethal effects would occur on a Phanerozoic Earth environment, but instead, a similar transient disturbance to the upper levels of the atmosphere as that imposed by only the electromagnetic radiation of solar flares or the recent galactic-scale GRB 221009A. These disturbances can be readily detected on Earth by very-low-frequency (VLF) amplitude measurements of the ionosphere, as well as the Geostationary Operational Environmental Satellite (GOES) X-ray sensor utilized for solar flares \citep{Hayes2017,Hayes2021}. Relatedly, the disturbances may offer interesting observational prospects for the examination of exoplanetary atmospheric compositions and processes \citep{Chen21}. Therefore, should an X-ray-luminous SN occur in the Milky Way, multiple pathways exist for modern instrumentation to measure these atmospheric disturbances.

A Milky Way SN has not been convincingly observed by the naked eye since Kepler’s SN 1604 and thus has not been detected by modern astronomical instruments. Interestingly however, there have been five SNe recorded in the historical record (seen on Earth by the naked eye) within the last millennium, with distance estimates ranging between 1.4 and 10 kpc. If but one of these were characteristic of an interacting SN, the possibility remains that their X-ray phase of emission caused a significant sudden ionospheric disturbance in Earth’s recent past.
\newpage

\section{Summary and Conclusions}
\label{sect:summ_and_conc}

In this paper, we have adopted the parameters of previous nearby SN research to conduct a threat assessment of the modern SN X-ray emission data of 31 SNe. We have restricted our initial data analysis to the empirically confirmed observations and have focused the majority of our initial threat assessment on the biological consequences related to ozone depletion. From these efforts alone, we have shown that the hazardous reach of these types of SNe is substantial.

Even in our most conservative estimates, the results show that SN X-ray emission has general implications for planetary habitability and potentially the evolution of Earth itself:

\begin{enumerate}

\item 
These events, while rare, maintain a notable influence in the radiation environment of the galaxy and pose a substantial threat to terrestrial biospheres, as their ionizing radiation can induce significant alterations to a planet’s atmospheric chemistry at formidable distances. We maintain strict standards for our calculations and analysis, adopting a high critical fluence value for the given energy ranges of the X-ray data and only considering the limited window of confirmed X-ray observations. Even with these restrictions, we have calculated that these SNe are capable of imposing lethal effects at distances well over the 10 pc standard of previous research.

\item
SN X-ray emission occurs as a distinct stage of an SN’s radiation emission for nearby planets within tens of parsecs: typically months/years after the initial outburst, and thousands of years before the arrival of cosmic rays. Therefore, a corollary of the formidable threat found here is that this alters the timeline by which we know an SN can influence a nearby planet, adding an additional phase of adverse effects.

\item
This lethality of X-ray-luminous SNe poses further constraints with respect to if and/or how life can evolve elsewhere in our galaxy and other star-forming regions. As we continue to detect more exoplanets and further the search for extraterrestrial life, SN X-ray emission needs to be considered in attempts to quantify habitability and/or locate potential biospheres. 

\item
We now open the discussion as to whether these X-ray-luminous SNe may have influenced life on Earth itself. The confirmed detection of SN radioisotopic material dating from the last $\sim$8 Myr of the geological record \citep{Wallner16} is consistent with the presence of our solar system within the Local Bubble, a hot, low-density region of space that is thought to be a product of numerous nearby SN explosions coinciding with Earth’s early Neogene Period ($\lesssim$ 20 Myr) \citep{Breit2016, Zucker22}. This means that a nearby SN has most certainly occurred in Earth’s geological past, likely numerous times. Combining these findings with our threat assessment here, it is possible that one or more of these SNe were interacting and thus inflicted a high dosage of X-ray radiation on Earth’s atmosphere. This would imply that SN X-ray emission has had a notable impact on Earth and potentially played a role in the evolution of life itself. 

\end{enumerate}

Here we have shown that simply from confirmed X-ray observations alone, the interacting X-ray phase of an SN’s evolution can entail significant consequences for terrestrial planets. We limit any further speculation until further developments in X-ray astronomy are made; however, the evidence presented here certainly points to this process being capable of imposing lethal consequences for life at formidable distances.

We thus conclude with the comment that further research into SN X-ray emission has value not just for stellar astrophysics, but also for astrobiology, paleontology, and the Earth and planetary sciences as a whole. We urge follow-up X-ray observations of interacting SNe for months and years after the explosion and urge for the continued development of X-ray telescopic instrument implementation in the hard X-ray band. These observations and innovations will shed light on the physical nature of SN X-ray emission and will clarify the danger that these events pose for life in our galaxy and other star-forming regions.

We gratefully acknowledge extensive illuminating discussions with Rafaella Margutti on X-ray supernovae. The work of I.R.B., C.M.O., and B.D.F. was supported in part by the NSF under grant No. AST-2108589.

\noindent
\software{
 Matplotlib \citep[][ http://dx.doi.org/10.1109/MCSE.2007.55]{matplotlib}, Numpy \citep{harris2020array}, Astropy\citep[][http://www.astropy.org]{astropy:2013, astropy:2018, astropy:2022}}
\newpage

\bibliography{XraySN}
\bibliographystyle{aasjournal}

\end{document}